\documentclass[12pt]{iopart}

\usepackage{iopams}  

\usepackage{graphicx}% Include figure files
\usepackage{amscd,amssymb}
\newcommand{\bm}[1]{\mbox{\boldmath $#1$}}
\usepackage{color}
\newcommand{\be}{\begin{equation}}
\newcommand{\bel}{\begin{equation}\label}
\newcommand{\ee}{\end{equation}}
\newcommand{\bc}{\begin{center}}
\newcommand{\ec}{\end{center}}

\begin{document}

\title[Self-similar solutions to the hypoviscous Burgers and SQG equations at criticality]
{Self-similar solutions to the hypoviscous Burgers and SQG equations at criticality}

\author{Koji Ohkitani}

\address{Research Institute for Mathematical Sciences,\\
  Kyoto University, Kyoto 606-8502 Japan.}

\ead{ohkitani@kurims.kyoto-u.ac.jp}

\begin{abstract}
  After reviewing the source-type solution of the Burgers equation with standard
dissipativity, we study the hypoviscous counterpart of the Burgers equation.
1) We determine an equation that governs the near-identity transformation underlying
its self-similar solution. 2) We develop its approximation scheme and construct the
first-order approximation. 3) We obtain the source-type solution numerically by the Newton-Raphson
iteration scheme and find it to agree well with the first-order approximation.
Implications of the source-type solution are given, regarding the possibility of linearisation
of the hypoviscous Burgers equation.
Finally we address the problems of the incompressible fluid equations in two dimensions,
centering on the surface quasi-geostrophic equation with standard and hypoviscous dissipativity.\\

{\it This Accepted Manuscript is available for reuse under a CC BY-NC-ND licence after the 12 month embargo period
  provided that all the terms and conditions of the licence are adhered to.}

\end{abstract}

\noindent{\it Keywords}: self-similarity, scale-invariance, hypoviscosity, Burgers equation, SQG equation 
%\maketitle

\section{Introduction}
%  Complete integrability
%  Elementary excitations in the decaying process
\color{black}

This work is motivated both mathematically and physically.
Mathematically,
in this paper we will be interested in forward self-similar solutions of fluid dynamical equations
that describe the form of flow fields in the late stage of decay.
This is conveniently represented by a self-similar profile that emerges after scaling transformation.

Self-similarity is a useful concept  when we handle linear and non-linear PDEs.
For linear PDEs with the Laplacian dissipative operator, the heat flow (i.e. the
solution to the heat equation) plays a central role. For example, even for non-linear PDEs, such as
the Navier-Stokes equations, their solutions eventually follow the heat flow when they exist for a long
time, e.g. \cite{OT2000}.

It is important to distinguish integrable systems from non-integrable ones.
%It is of mathematical interest to seek a transformation that linearises, albeit approximately,
%the fully nonlinear equation.
In some cases the profiles of partial differential equations (PDEs, hereafter) coincide with those of
the linear equation (e.g. the heat equation). In other cases they don't where the trace of nonlinear term
remains in the profile and it may give a clue as to how to
study the PDE throughout the whole time evolution. This is particularly the case when we study
integrable equations that are soluble as a function of heat solutions. 
This is because a trace of nonlinear terms, if it is present in the self-similar profile, 'inherits'
the information about the function required to linearise the fully nonlinear problem.
Note that the linearising function is independent of time by definition.
Hence once it is determined it can also cover the earlier stage when the nonlinearity is comparable to dissipation.

Physically we consider the SQG equation with the hypoviscous dissipation (particularly
the square-root of the Laplacian) corresponding to the physical mechanism of Ekman pumping,
e.g. \cite{Constantin2002}. This case is critical in the sense that dissipation and nonlinearity
balances with respect to conservation laws and it is known to be globally well-posed \cite{KNV2007}.
There are a range of one-dimensional counterparts for the SQG equation \cite{CCCF2005,Do2014}, which has
nonlocal advecting velocity. As a simpler equation we may consider a model with local velocity,
the  hypoviscous version of the Burgers equation,
which is known to be critical and globally well-posed \cite{KNV2007}.
While the boarder-line case is well-posed, we do not know whether it is integrable via the 'heat' kernel method
just like the standard Burgers equation is.
%It seems worth studying how to solve non-linear PDEs on the basis of self-similar solutions.
In this paper we consider the  hypoviscous  Burgers equation,
raising the questions as to (i) whether it is integrable or not and (ii) if so, how we may in principle achieve
that.
We also consider self-similar solutions of the  hypoviscous version of the SQG
equation, which has more physical significance.

This work may be regarded as a testbed upon which we illustrate and expand the ideas for studying more
realistic equations, including the 3D Navier-Stokes equations, e.g. \cite{OV2022}.
However, the results are of interest on its own theoretical footing.
The main tools we will employ are dynamic scaling transformations and critical scale-invariance.

\color{black}
The rest of this paper is organised as follows. In Section 2 we summarise the analyses of the Burgers equation
with standard Newtonian dissipativity. While this part is basically a survey, we present probably the
simplest possible derivation
of the Cole-Hopf transformation by writing down the {\it determining  equation} for the near-identity
transformation underlying such a linearisation. 
We also demonstrate how the expression of the  exact solution
(equivalent to the so-called Cole-Hopf linearisation) is retrieved through a chain of approximations and asymptotics.
In Section 3 we consider the hypoviscous Burgers equation and derive the determining equation
for its self-similar solution and present its approximate solutions by an asymptotic analysis.
The self-similar profile is determined by solving the fully-nonlinear equation using
the Newton-Raphson iteration method and compared with the approximate solution.
We also consider how we can construct the more general solutions in the original variables.
In Section 4 we consider radially symmetric self-similar solution of the SQG equations
with standard and hypoviscous dissipativity. Section 5 is devoted to a summary and outlook.

\section{Burgers equation with standard dissipativity}
\subsection{Dynamic scaling} 
We consider the Burgers equation in $\mathbb{R}^1$
\bel{Burgers}
\frac{\partial u}{\partial t} + u \frac{\partial u}{\partial x}
=\nu \frac{\partial^2 u}{\partial x^2},
\ee
under the boundary condition $u(x,0)=u_0(x),\, |u(x,t)| \to 0 \;\mbox{as}\;|x| \to \infty,$
where $u_0$ denotes an initial data.
We  review the method of the linearisation of the Burgers equation stressing and recapitulating the
 basic concepts underlying its solutions.
It is well-known that the so-called Cole-Hopf transformation \cite{Hopf1950, Cole1951}
\bel{C-H}
u(x,t)=-2\nu \frac{\partial}{\partial x} \log \theta(x,t)
\ee
reduces it to the heat equation
\be
\frac{\partial\theta}{\partial t}=\nu \frac{\partial^2\theta}{\partial x^2},
\ee
which is soluble by the heat kernel
$$\theta(x,t)=   \frac{1}{\sqrt{4\pi \nu t}}\int_{-\infty}^{\infty} \exp \left( -\frac{(x-y)^2}{4\nu t}\right) \theta_0(y) dy.$$
The Hopf's derivation of the transform is concise and succinct where the final form is given as is without
elucidating any reason.
On the other hand, the Cole derived it  relying explicitly on the concept
of scale-invariance,
urging us to go for the velocity potential $\phi$.
Indeed, Cole suspected that we might be able to write $\phi(x,t)=F(\theta(x,t))$ for some function $F$,  and he was able to
determine $F$ successfully.
  
Cole's argument can be summarised as follows. The equation (\ref{Burgers}) satisfies the following scaling laws
$$ x \to \lambda x, t \to \lambda^2 t, u \to \lambda^{-1} u,\;\;\mbox{for}\;\forall \lambda >0.$$
that is, if $u(x,t)$ is a solution, so is $u_\lambda(x,t) \equiv \lambda u(\lambda x, \lambda^2 t).$
If we go for the velocity potential $\phi,$ defined by $u=\frac{\partial \phi}{\partial x},$  its governing equation takes
the following form of the Hamilton-Jacobi equation
$$  \frac{\partial \phi}{\partial t}+ \frac{1}{2}\left(\frac{\partial \phi}{\partial x}\right)^2
=\nu \frac{\partial^2 \phi}{\partial x^2}.$$
Here, the scaling laws are altered accordingly, as
$x \to \lambda x, t \to \lambda^2 t, \phi \to \lambda^{0} \phi;$ that is, if $\phi(x,t)$ is a solution,
so is $\phi_\lambda(x,t) \equiv \lambda^{0}  \phi(\lambda x, \lambda^2 t).$ 
We call this fact {\it type 1 critical scale-invariance}, which can be traced back to the fact that
the physical dimension of $\phi$ is the same as that of $\nu$, that is, $[\nu]=\frac{L^2}{T},$ where
$[\cdot]$ denotes the physical dimensions and $L,T$ length and time.

To handle the problem systematically we introduce  dynamic scaling transforms:
$$U(\xi,\tau)=\lambda(t)u(x,t),\;\lambda(t)=\sqrt{2 a(t+t_*)},$$
$$\xi=\frac{x}{\lambda(t)},\;\tau=\frac{1}{2 a}\log \frac{t+t_*}{t_*},$$
where we take $2at_*=1$ to initialise $u(x,0)=U(\xi,0)$.
\textcolor{black}{Dynamic scaling transforms have been used in many publications, readers are recommended to
  consult   e.g. \cite{GGS2010, Chae2007, OV2022}.}
Here the zooming-in parameter $a$ is on the same order of $\nu$.
The Burgers equation after the transformations becomes
  $$\frac{\partial U}{\partial \tau}
+U \frac{\partial U}{\partial \xi}
=\nu \frac{\partial^2 U}{\partial \xi^2}+
\underbrace{ a \xi \frac{\partial U}{\partial \xi}  + aU}_{
  =a  \frac{\partial}{\partial \xi}\left( \xi U \right)},$$
where the additional terms complete a divergence form (i.e. a total derivative).
We call this fact {\it type 2 critical scale-invariance}, \textcolor{black}{under which the steady linear solution
is given by the heat kernel  (in the case of Laplacian dissipativity).}
Incidentally, in term of  $\phi(x,t)=\Phi(\xi,\tau),$ with $\Phi(\xi,\tau)$ defined by
$U(\xi,\tau)=\partial_{\xi}\Phi(\xi,\tau)$ the scaled equation reads
$$\frac{\partial \Phi}{\partial \tau}
+ \frac{1}{2}\left(\frac{\partial \Phi}{\partial \xi}\right)^2
=\nu \frac{\partial^2 \Phi}{\partial \xi^2}+a  \xi\frac{\partial \Phi}{\partial \xi}.$$
Observe that in velocity potential the  the dynamically scaled equation 
and the original one are apart only by the drift term.

\subsection{Cole's argument {\it after dynamic scaling}}
We will apply Cole's argument to the dynamically-scaled equation to demonstrate that
it is an expedient  of recovering the Cole-Hopf linearisation. That way we can put
the more difficult problem of
time evolution on the back burner and focus our effort in solving a steady nonlinear equation
that is more tractable. 

Assume that $\Phi(\xi)=Cf(s)$ for some function $f$\footnote{In this paper $C$ takes different values
  from place to place.}, where
$s \equiv\int_0^\xi \exp \left(-\frac{a \eta^2}{2\nu}\right) d\eta$ denotes
the scaled heat flow and  $C$ a constant. \textcolor{black}{This is the key assumption which works fine for integrable
  systems; if a (time-independent) near-identity is recoverable by examining the late stage, it can cover the
  whole time evolution. Generally speaking, such an approach would not work for non-integrable systems, at least as is.} 
  
We then have $U=\Phi_\xi=C \exp \left(-\frac{a \xi^2}{2\nu}\right) f'(s),$
from which it follows that
$$U_\xi+\frac{a}{\nu}\xi U=C \exp \left(-\frac{a \xi^2}{\nu}\right) f''.$$
Plugging these into the steady version of the scaled equation
$$\frac{U^2}{2}=\nu U_\xi+ a\xi U$$
we obtain
\bel{deteq_standard}
\frac{dg}{ds}=\frac{R}{2} g^2,
\ee
where we have put $g\equiv f'$
\textcolor{black}{and $R \equiv\frac{C}{\nu}$ denotes the Reynolds number.}
Here $C$ is a constant that depends on $M=\int_{-\infty}^{\infty} U(\xi) d\xi.$
This innocent-looking equation (\ref{deteq_standard}) contains vital information as to how we may solve equation (\ref{Burgers}).
Indeed it determines the transformation
$$g(s)=\frac{1}{1 -\frac{R}{2}s}.$$
Equivalently we have in terms of $f$
$$ f(s)=-\frac{2}{R}\log\left(1 -\frac{R}{2}s \right).$$
and $f$ is a near-identity transformation in the sense that $f(s) \approx  s,\;\;\mbox{for}\;\; R \ll 1.$

%$$U_{\xi \xi} +\frac{a}{\nu}(\xi U)_\xi =\frac{1}{\nu}\left( \frac{U^2}{2}\right)_\xi$$
In terms of $U$, the solution takes the well-known near-Gaussian form
\bel{Burgers_s-s}
U(\xi)=\frac{C \displaystyle{\exp \left( -\frac{a \xi^2}{2 \nu} \right)}}
  {1  -\displaystyle{\frac{R}{2}\int_0^{\xi}}
    \exp \left( -\frac{a \eta^2}{2 \nu}\right) d\eta},
  \ee
see  e.g. \cite{LP1984, EZ1991, BKW1999}.
Recasting the above, we find
  $$U(\xi)=- 2 \nu \partial_\xi \log \left( 1-\frac{R}{2} \int_0^{\xi}
\exp \left( -\frac{a \eta^2}{2 \nu}\right) d\eta \right),$$
which is already the Cole-Hopf transform in disguise.
The constant $C$ turns out to be given explicitly by
$C=\sqrt{\frac{8a\nu}{\pi}}\tanh \left(\frac{M}{4\nu}\right)$
and satisfies  $C\approx \sqrt{\frac{a}{2\pi \nu}}M,\;\;\mbox{if}\;\;
\frac{M}{\nu} \ll 1.$

With one more  step, called a lifting procedure,  we can recover
the full expression for general solutions, as recalled in Section 2.4 below.
It is to be noted that small deviation from the Gaussian can tell us how to handle more general solutions.
In other words, critical information is encoded in the tiny part as to how we may linearise the fully nonlinear
equation.

For the standard dissipativity of the Laplacian form, 
after long time evolution remnants of the nonlinear terms remain in the self-similar profile,
that is, the Laplacian cannot eradicate  nonlinearity completely.
With weaker dissipativity of the Zygmund operator, {\it a fortiori} there should be a function class
(under suitable boundary conditions) where nonlinearity survives. That is the situation we are after in this paper .

\subsection{Poincar{\'e}'s variational equation}

\textcolor{black}{A perturbative treatment  for handling differential equation with a parameter is known
  as  Poincar{\'e}'s method, see e.g. \cite{Yosida1960}.
As an illustration we describe a method of solving (\ref{deteq_standard}) approximately, 
the determining equation for the near-identity where the unperturbed solution is the one for $R=0.$}
\footnote{It is possible  to eliminate $R$ by the replacement \textcolor{black}{$s \to s/R.$}
    However, we don't do that here
  because we intend to carry out a perturbative argument for small $R$.}
At the unperturbed state $R=0$, we have
$$\frac{dg}{ds}=0,$$
which can be solved trivially as $g={\rm const.,}$ \textcolor{black}{
but this const. should be taken to be unity, as we go for a near-identity $f(s) \approx s$.}
Next consider $z \equiv \frac{\partial g}{\partial R}$, the variation with respect to $R$, and we write
its equation as
$$\frac{d z}{ds}=\Bigl. \left(\frac{1}{2}g^2 +R g z\right) \Bigr|_{R=0}=\frac{1}{2}.$$
\textcolor{black}{
Its solution is  $z=s/2+z(0)$,  but we can take the 'initial condition' $z(0)=0$ because  
$s=\widehat{\Psi}$ represents the velocity potential and an additive constant is insignificant.}
Thus we find, to the first-order,
\begin{eqnarray}
  g(s) &=& 1+R\frac{s}{2} \nonumber \\
  &\approx& \frac{1}{1-\frac{R}{2}s},\nonumber
\end{eqnarray}  
  where  rationalisation has been introduced in the final line under the assumption $R \ll 1$,
  somewhat in the spirit of Pad{\'e} approximation. It should be noted that this retrieves
  the exact self-similar solution  of the Burgers equation.

\subsection{Lifting the self-similar solution to a more general one}
For convenience we recapitulate the procedures of lifting  described in \cite{KO2022}.
We assume that the Reynolds number $R$ is small.\\
  {\bf Step 1.} Assume $\Psi(\xi)=C f(s),
  s=\widehat{\Psi}=\int_0^{\xi}\exp\left(-\frac{a\eta^2}{2\nu} \right)$ and we write
\begin{eqnarray}
   U(\xi) &=& C f'(s) e^{-\frac{a\xi^2}{2\nu}} \nonumber \\
  &=&  F( \widehat{\Psi}(\xi);C\partial_\xi \widehat{\Psi}(\xi))\nonumber \\
  &\equiv& \frac{C\partial_\xi \widehat{\Psi}}{1-\frac{C}{2\nu}\widehat{\Psi}}.\nonumber
\end{eqnarray}  
Here the function $F$ satisfies the scaling $F(x;\alpha y)=\alpha F(x;y)$ for $\forall \alpha >0$
and $U(\xi)$ is a \textit{near-identity} transformation of $C\partial_\xi \widehat{\Psi}(\xi),$
more precisely we have
$$F(x;y)=\frac{y}{1-\frac{C}{2\nu}x},$$
where $C/\nu$ is small.
Reverting to the original variables by definition, a particular self-similar solution is obtained as
\begin{eqnarray}
u(x,t)&=&\frac{1}{\sqrt{2a(t+t_*)}}F\left(\widehat{\Psi}\left(\frac{x}{\sqrt{2a(t+t_*)}}\right);C\partial_\xi
\widehat{\Psi}\left(\frac{x}{\sqrt{2a(t+t_*)}}\right)\right)\nonumber \\
&=& \frac{1}{\sqrt{2a(t+t_*)}}\frac{C\partial_\xi
\widehat{\Psi}\left(\frac{x}{\sqrt{2a(t+t_*)}}\right)}{1-\frac{C}{2\nu}\widehat{\Psi}\left(\frac{x}{\sqrt{2a(t+t_*)}}\right)}.
\nonumber
\end{eqnarray}
We now generalise the above expression to cover arbitrarily large Reynolds number.\\
{\bf Step 2.} Replacing the self-similar heat flow with the general  one, we have
\begin{eqnarray}
u(x,t)&=&\frac{1}{\sqrt{2a(t+t_*)}}
 F\left(\widehat{\psi}\left(x,t\right); C\sqrt{2a(t+t_*)}\partial_x \widehat{\psi}\left(x,t\right)\right) \nonumber\\
 &=&C F\left(\widehat{\psi}\left(x,t\right);\partial_x \widehat{\psi}\left(x,t\right)\right)\\
 &=& \frac{C\partial_x \widehat{\psi}(x,t)}{1-\frac{C}{2\nu} \widehat{\psi}(x,t)},\nonumber
\end{eqnarray}
where
$$
\widehat{\psi}(x,t)=\frac{1}{\sqrt{4\pi\nu t}}\int_{-\infty}^{\infty} \widehat{\psi}(y,0) \exp\left(-\frac{(x-y)^2}{4\nu t} \right)dy
$$
denotes the general solution to the linearised equation  by the heat kernel. Hence we find
\begin{eqnarray} 
u(x,t)
&=& -2\nu \partial_x \log\left|1-\frac{R}{2}\widehat{\psi} \right|\nonumber \\
&=& -2\nu \partial_x \log\left|\widehat{\psi}-\frac{2}{R} \right|.\nonumber  
\end{eqnarray}
When $R \ll 1,$ we have $u(x,t) \approx C\partial_x \widehat{\psi}$ by the first expression,
whereas $R \gg 1,$ $u(x,t) \approx -2\nu \partial_x  \log \widehat{\psi}$  by the second.
In the above lifting, the virtual time origin $t=-t_*$ is shifted to $t=0$, for simplicity.

{\bf Remark}
The velocity $U$ can also be expressed solely in terms of
$\widehat{U}(\xi) \equiv C \exp \left( -\frac{a \xi^2}{2 \nu} \right)$ as
$$U(\xi)=\frac{\widehat{U}(\xi)}{1\mp\sqrt{\frac{\pi}{8a\nu}}{\rm erf} \left( \sqrt{\left| \log\frac{\widehat{U}(\xi)}{C} \right|}\right) },$$
which is {\it not} revealing. This is because
whilst the dominant contribution (the numerator) is translated appropriately when we make a replacement
$\widehat{U}(\xi) \to \sqrt{2a(t+t_*)}\partial_x \widehat{\psi}(x,t),$ but the sub-dominant contribution (the denominator) is {\it not}, as can be gathered from
$$\sqrt{\frac{\pi}{8a\nu}}{\rm erf} \left( \sqrt{\left| \log\frac{\sqrt{2a(t+t_*)}\partial_x \widehat{\psi}(x,t)}{C} \right|}\right)
\ne\frac{C}{2\nu} \widehat{\psi}(x,t).$$
This shows the importance of choosing the right variable; when we express the nonlinear solution 
by the linear solution we should choose the dependent variable so as to satisfy the scale invariance of type 1.
In the case of the Laplacian dissipativity, the velocity potential is the right answer.

\subsection{Determination of the self-similar profile by the Newton-Raphson method}
As a preparation for handling the hypoviscous counterpart,
we consider the scaled Burgers equation and study its steady solution numerically
\textcolor{black}{using finite differences, see Appendix A.1 for details}. 
We pretend that we simulate it in $\mathbb{R}^1$ by solving it on a large,
but finite spatial interval for flows that are well-localised.
We need to check whether the method actually works or not, using a problem
for which an exact solution  is available.

Integrating \textcolor{black}{the steady version of the scaled equation}
\bel{steady_sc_Burg}
\nu U_{\xi\xi}+a(\xi U)_{\xi} -U U_{\xi}=0
\ee
and drooping the integration constant \textcolor{black}{under the boundary conditions
$U(\xi) \to 0$ as $|\xi| \to \infty$}, we find
$$U^2=2(\nu U_{\xi}+a\xi U).$$

We take the initial guess as the linear solution, i.e. the Gaussian form
$$U^{(0)}=\sqrt{\frac{a}{2\pi\nu}}\exp\left(-\frac{a\xi^2}{2\nu}\right).$$
The numerical parameters used are $\nu=a=1, L=0, M \approx 2.633$  and  thus $C=\frac{8}{\sqrt{\pi}}\tanh \frac{M}{4}$
in (\ref{Burgers_s-s}).
We carried out iterations by using the $L$-$U$ decomposition. An excellent convergence is achieved just after 10 iterations.

\begin{figure}
  \begin{minipage}{.7\linewidth}\
  \includegraphics[scale=.7,angle=0]{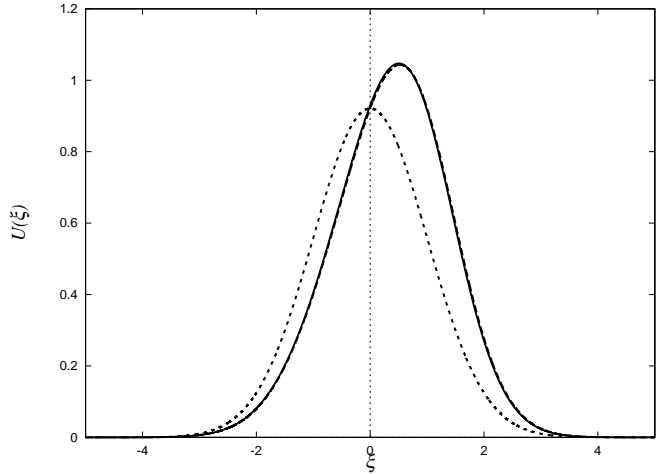}
  \caption{Self-similar profile $U(\xi)$ of the standard Burgers equation in velocity;
    the exact solution (solid), the first-order approximation (dashed). The zeroth-order
    approximation (dotted) is also shown for comparison.}
  \label{Fig.1}
  \end{minipage}
\end{figure}
In Fig.\ref{Fig.1} we see that
the solution obtained by the Newton-Raphson method (solid) matches perfectly with the exact solution (dashed).
  The linear solution (dotted) is also shown for comparison. The deviation \textcolor{black}{from the linear solution} is  not large, but appreciable.

  \section{Hypoviscous Burgers equation}
\subsection{Governing equations}
  We  consider the following Burgers equation with hypoviscous dissipativity
  \bel{hypoBurgers}
  \frac{\partial u}{\partial t} + u \frac{\partial u}{\partial x}=-\nu' \Lambda u,
  \ee
  where  $\Lambda \equiv (-\partial_{xx})^{1/2}=\partial_x H$ denotes the Zygmund operator  and
  $$H\left[f\right]=\frac{1}{\pi}{\rm p.v.}\int_{-\infty}^{\infty}\frac{f(y)}{x-y}dy$$ the Hilbert transform, defined with
  a principal value integral.
    Well-posedness of (\ref{hypoBurgers}) has been established in  \cite{KNS2008}
    for more general form of dissipativity $-\nu' \Lambda^\alpha u$ where $1/2 \leq \alpha \leq 1.$
    We will focus on the case $\alpha=1/2,$ for which an integral expression of the dissipativity
    is given explicitly  by the Poisson kernel. \textcolor{black}{It should be noted that only in this case
      can we carry out detailed analyses  because formulas of the Hilbert transforms are useful for them,
      which is not the case for $1/2 < \alpha <1$. }
    
 For (\ref{hypoBurgers}), the scale invariance is altered as follows
    $$ x \to \lambda x, t \to \lambda t, u \to \lambda^{0} u$$
    to the effect that if $u(x,t)$ is a solution, so is $u_\lambda(x,t) \equiv \lambda^{0} u(\lambda x, \lambda t).$
    Note that {\it type 1 critical scale-invariance} is achieved with the velocity because of physical dimensions
    $[u]= [\nu']=\frac{L}{T}.$
The governing equation for the velocity gradient $w(x,t)$ reads    
$$\frac{\partial w}{\partial t}+u\frac{\partial w}{\partial x}=-w^2 -\nu' \Lambda w,$$
whose scaling law is given by
$$ x \to \lambda x, t \to \lambda t, w \to \lambda^{-1} w,$$
that is, if $w(x,t)$ is a solution, so is $w_\lambda(x,t) \equiv \lambda w(\lambda x, \lambda t).$

When the nonlinear term is discarded in (\ref{hypoBurgers}),
the linear solution can be obtained by using the Poisson kernel
\bel{Poisson}
u(x,t)=\frac{\nu' t}{\pi}\int_{-\infty}^{\infty} \frac{u_0(y)dy}{(x-y)^2+(\nu' t)^2},
\ee
which can be regarded as an analogue of the heat kernel for the standard Burgers equation.

\subsection{Dynamic scaling}
Under the dynamic scaling transformations
$$U\left( \xi,\tau \right)=u(x,t),
\xi=\frac{x}{\lambda(t)}, \tau=\frac{1}{a}\log \frac{t+t_*}{t_*},\;\; \lambda(t)=a(t+t_*),$$
where $at_*=1,$
the equation (\ref{hypoBurgers}) is transformed to
$$
\frac{\partial U}{\partial \tau}+U\frac{\partial U}{\partial \xi}=-\nu' \Lambda U+a \xi \frac{\partial U}{\partial \xi}.
$$
The governing equation for the  velocity gradient $W\left( \xi,\tau \right)=\lambda(t)w(x,t)$ reads
$$
\frac{\partial W}{\partial \tau}+U\frac{\partial W}{\partial \xi}=-W^2 -\nu' \Lambda W
+a\frac{\partial}{\partial \xi}(\xi W).
$$
Rewriting (\ref{Poisson}) in scaled variables, we find
\color{black}
\bel{U}
\widehat{U}(\xi,\tau)=\frac{\nu' }{\pi}\frac{1-e^{-a \tau}}{a}
\int_{-\infty}^{\infty} \frac{U_0(e^{a\tau}\eta)d\eta}{(\xi-\eta)^2+\left(\nu'\frac{1-e^{-a \tau}}{a} \right)^2}.
\ee
Note that on this basis it is virtually impossible to predict the long-time limit by passing to the limit $\tau \to \infty$.
By differentiating (\ref{U}), we find the linear solution $\widehat{W}$ in scaled form as
$$\widehat{W}(\xi,\tau)=\frac{\nu' }{\pi}\frac{1-e^{-a \tau}}{a}
\int_{-\infty}^{\infty} \frac{e^{a\tau}W_0(e^{a\tau}\eta)d\eta}{(\xi-\eta)^2+\left(\nu'\frac{1-e^{-a \tau}}{a} \right)^2},$$
\textcolor{black}{where $W_0 (> 0)$ denotes an initial data.}
It follows that
$$\lim_{\tau \to \infty} \widehat{W}(\xi,\tau)=\frac{M}{\pi}\frac{\mu}{\xi^2+\mu^2},$$
where  $\mu=\nu'/a,$ $M\equiv \int_{-\infty}^{\infty} w dx.$ Here we have made use of a property of
the Dirac delta function 
$$\lim_{\epsilon \to 0} \frac{1}{\epsilon}W_0\left( \frac{\eta}{\epsilon}\right) =M\delta(\eta),$$
\textcolor{black}{which holds for any localised initial data $W_0$.}
Thus, by $\widehat{U}(\xi)=\int_0^\xi \widehat{W}(\eta)d\eta$ we find retrospectively that
$$\lim_{\tau \to \infty} \widehat{U}(\xi,\tau)=\frac{M}{\pi}\tan^{-1}\frac{\xi}{\mu}.$$
It is also noted that a different  scaling was adopted for the analysis of
(\ref{hypoBurgers}) under a stronger boundary condition
$u \in L^1(\mathbb{R}^2)$ in \cite{Iwabuchi2015, Iwabuchi2020}.
\color{black}

We summarise self-similar solutions of the  Burgers equations with standard and hypoviscous
dissipativity in Table \ref{Burgers_ss}.

\begin{table}[h]                                                                
\caption{Suitably normalised self-similar solutions of Burgers equations}
\begin{center}   
\begin{tabular}{ccc}\hline                                                      
dissipativity  & Burgers equations \\ \hline
Laplacian operator $\triangle$ \\($\nu/a=2$) &  $\begin{array}{l}\mbox{type 1:}\; \Phi = \log(1-C\frac{\sqrt{\pi}}{2}\,{\rm erf}(\xi))\\
  \mbox{type 2:}\; :U = \frac{Ce^{-\xi^2}}{1- C\frac{\sqrt{\pi}}{2}\,{\rm erf}(\xi)}\end{array}$  \\\noalign{\vskip 0.2cm} \hline
Zygmund operator $\Lambda$\\($\nu'/a=1$)& $\begin{array}{l} \mbox{type 1:}\; U = \frac{C'}{\pi}\tan^{-1}\xi \\\mbox{type 2:}\; W = \frac{C'}{\xi^2+1} \end{array}$ \\ \noalign{\vskip 0.2cm}\hline 
\end{tabular}
\end{center}
\label{Burgers_ss}
\end{table}   

In order to study the self-similar profile $W(\xi)=\lim_{\tau \to \infty}W(\xi,\tau)$,
we set $W_\tau=0$ in the dynamically-scaled equation 
and consider the steady solution defined by
\bel{steadyNL1}
(UW)_\xi= -\nu' \Lambda_\xi W +a(\xi W)_\xi.
\ee
%$$UW= -\nu' H[W] +a\xi W,$$
Integrating (\ref{steadyNL1}) and dropping the constant of integration
under the boundary conditions $W(\xi) \to 0$ as $|\xi| \to \infty$, we find
\bel{steadyNL2}
    \xi W  -\mu H[W] =\frac{1}{a}UW.
\ee
We now begin studying (\ref{steadyNL2}). If the nonlinear term is discarded completely, we have 
$$\xi W  -\mu H[W] =0.$$
Recalling the basic properties of the Hilbert transform
$$H\left[\frac{\mu}{\xi^2+\mu^2} \right]=\frac{\xi}{\xi^2+\mu^2},\;
H\left[\frac{\xi}{\xi^2+\mu^2} \right]=-\frac{\mu}{\xi^2+\mu^2},$$
for $\mu >0,$ it is easily verified that
$$W \propto \frac{\mu}{\xi^2+\mu^2}$$
solves the linearised equation.

%$$U(\xi) \approx C \tan^{-1}\frac{\xi}{\mu},\;W(\xi) \approx C\frac{\mu}{\xi^2+\mu^2},\;\;C=M/\pi$$
  Assuming that  
  $$U(\xi)=C f(s)$$
  for some function $f$ and a constant $C,$ 
  where $s \equiv \tan^{-1}\frac{\xi}{\mu},$
  we  find
\bel{Weq}  
  W(\xi)=C\frac{\mu}{\xi^2+\mu^2} f'(s).
\ee
It follows from (\ref{steadyNL2}) that 
$$
  \frac{C}{a}\frac{ff'}{\xi^2+\mu^2}=\frac{\xi}{\xi^2+\mu^2}f'
  -H\left[\frac{\mu}{\xi^2+\mu^2} f' \right].
%&=&H\left[\frac{\mu}{\xi^2+\mu^2} \right]f'-H\left[\frac{\mu}{\xi^2+\mu^2} f' \right].
$$
  It is of interest to note that we can also write the above as
  $$\frac{C}{a}\frac{ff'}{\xi^2+\mu^2}=-[H,f']\frac{\mu}{\xi^2+\mu^2},$$
  where $[A,B]=AB-BA$ denotes a commutator. 
Those expressions are still mixed in that the independent variable $\xi$ and
$d/ds$ coexist therein. By $s=\tan^{-1}\frac{\xi}{\mu},$ we can convert them in a form where
only $s$ appears as the  independent variable.
To this end, we consider the Hilbert transform of any function $h(\cdot)$ 
\begin{eqnarray}
H[h]&=&\frac{1}{\pi}{\rm p.v.}\int_{-\infty}^{\infty}\frac{h(\eta)}{\xi-\eta}d\eta \nonumber\\
&=&\frac{1}{\pi}{\rm p.v.}\int_{-\pi/2}^{\pi/2}\frac{\bar{h}(r)}{\cos^2 r}\frac{d r}{\tan s-\tan r},\;\;
\mbox{by}\; r=\tan^{-1}\frac{\eta}{\mu}\nonumber\\
&=& K\left[ \frac{\bar{h}(s)}{\cos^2 s}\right], \nonumber
\end{eqnarray}
where $h(\xi)=h(\mu \tan s)\equiv \bar{h}(s)$ and
$$K[g]\equiv\frac{1}{\pi}{\rm p.v.}\int_{-\pi/2}^{\pi/2}\frac{g(r)dr}{\tan s -\tan r}$$
denotes a modified version of the Hilbert transform.

It is in order to state some properties of $K[\cdot]$.
Taking $\bar{h}(s)=1,$ we get
$$K\left[ \frac{1}{\cos^2 s}\right]=H[1]=0.$$
Also, taking $\bar{h}(s)=\cos^2 s=\frac{\mu^2}{\xi^2+\mu^2}=f(\xi)$ and noting
$H[f]=\mu\frac{\xi}{\xi^2+\mu^2},$ we have $$K[1]=\cos s \sin s.$$
 After straightforward rearrangements we find
\bel{deteq_hypo}
R ff'\cos^2 s =K[1]f'-K[f'],
\ee  
where $R=\frac{C}{\nu'}.$ Note that the right-hand side can be identified as a commutator $-[K,f']1.$

The equation (\ref{deteq_hypo}) is the determining equation for the near-identity transformation $f(s)$,
which should be contrasted with the determining equation (\ref{deteq_standard})
for the  standard Burgers equation.

The determining equation (\ref{deteq_hypo}) for the near-identity transformation can be recast as
\bel{deteq_hypo2}
f'-\frac{1}{\cos s \sin s}K[f']=R ff' \cot s,
\ee
where  $R\ll 1$. It should be noted that no approximations have been introduced so far.

The fact that we can obtain an equation for the near-identity transformation $f(s)$
solely in terms of $s$ indicates that the hypoviscous Burgers equation is
integrable by the Poisson kernel. Indeed, using a standard perturbative argument we can
show that a solution $f(s)$ exists in the neighborhood 
of small $R$ on the basis of the existence of the linear solution, i.e. $f'(s)=1$.
The proof is similar to that of the existence of solutions to an ODE with a parameter,
e.g. \cite{Yosida1960}.

\subsection{Formal solution of the variational equation via a Neumann series}
While the existence of a solution to (\ref{deteq_hypo2}) can be shown on the basis of
a perturbative argument around $f(s)=s$ at $R=0$ \cite{Yosida1960},
it is not known how to solve it exactly.
We therefore seek its approximate solution that is valid for small $R$.

Consider the variation $z=\frac{\partial f}{\partial R}$ with respect to $R$, then the governing equation for $z$ reads
$$z'-\frac{1}{\cos s \sin s}K[z']=\Bigl. ff'\cot s\Bigr|_{R=0} +\Bigl.R (z f'+fz')\cot s\Bigr|_{R=0},$$
that is,
$$ \left(1-\frac{1}{\cos s \sin s}K[\cdot]\right) z'(s)=s\cot s,$$
because $f(s)=s$ at $R=0$.
Formally inverting the integral operator, we find
  \begin{eqnarray}
  z'(s)&=&\left(1-\frac{1}{\cos s \sin s}K[\cdot]\right)^{-1} s\cot s \nonumber \\
  &=&\sum_{n=0}^\infty \left(\frac{1}{\cos s \sin s}K[\cdot]\right)^{n} s\cot s \nonumber \\
  &=&s \cot s+ \frac{1}{\cos s \sin s}K[s\cot s]+\ldots,\nonumber 
  \end{eqnarray}
  by Taylor-expanding the right-hand side in the Neumann series.

  The near-identity transformation can be reconstructed, to the first-order in $R$, as
  $$ f(s) = s  +R z(s),$$
  and \textcolor{black}{retaining the first term in $z'(s)$},
  we first study  the leading-oder approximation of its derivative $g=f'$
  $$g(s) = 1+R z'(s)$$  
  $$ \approx 1+ R s\cot s.$$  
The first-order approximation to  (\ref{deteq_hypo2}) that retains only one term of
the Neumann series for the principal variational equation 
is thus given by
$$W(\xi) =C(1+R s \cot s)\frac{\mu}{\xi^2+\mu^2}$$
and its prefactor $C$ can be fixed as follows
$$M=\int_{-\infty}^{\infty} W(\xi) d\xi=C\left( \pi +R \int_{-\infty}^{\infty} s \cot s \frac{\mu}{\xi^2+\mu^2} d\xi\right).$$
Because the above definite integral is evaluated as $\int_{-\pi/2}^{\pi/2} s \cot s ds=\pi \log 2,$
we find $M=C\pi(1+R \log 2).$ The first-order approximation thus reads
\bel{W_1st}
W(\xi)=\frac{M}{\pi}\frac{1}{1+R \log 2}\left( 1+R \frac{\tan^{-1}\xi/\mu}{\xi/\mu}\right) \frac{\mu}{\xi^2+\mu^2}.
\ee
Below we will check the performance of the approximation
by comparing the source-type solution obtained by the Newton-Raphson method.

In order to find the corresponding $f(s)=s+R I(s)$, it is necessary to evaluate the indefinite integral
$I(s)=\int_0^s r\cot r\, dr.$
Actually, we show as a Lemma 1 in Appendix B 
$$I(s)=s\log(2\sin s)+2\sum_{n=1}^{\infty}\frac{\sin(2ns)}{(2n)^2}.$$

In passing, we will take a brief look at the second-order term in the Neumann series for $z'(s)$.
By the  product formula for the Hilbert transform
\textcolor{black}{
$$H[pq]=H[p]q+pH[q]+H[H[p]H[q]],$$
which holds for any functions $p,q$,}
we have
\begin{eqnarray}
H\left[ \frac{\mu}{\xi} \frac{\tan^{-1}\frac{\xi}{\mu}}{\xi^2+\mu^2}\right]
&=&H\left[ \frac{\mu}{\xi} \right]\frac{\tan^{-1}\frac{\xi}{\mu}}{\xi^2+\mu^2}
+ \frac{\mu}{\xi} H\left[ \frac{\tan^{-1}\frac{\xi}{\mu}}{\xi^2+\mu^2}\right]
+H\left[H \left[\frac{\mu}{\xi}\right] H\left[\frac{\tan^{-1}\frac{\xi}{\mu}}{\xi^2+\mu^2}\right]\right] \nonumber\\
&=&\frac{\mu}{\xi}\frac{1}{\xi^2+\mu^2}\log\frac{\sqrt{\xi^2+\mu^2}}{2\mu}
+H\left[ -\mu \pi \delta(\xi) \frac{1}{\xi^2+\mu^2}\log\frac{\sqrt{\xi^2+\mu^2}}{2\mu}\right] \nonumber\\
&=&\frac{\mu}{\xi}\frac{1}{\xi^2+\mu^2}\log\frac{\sqrt{\xi^2+\mu^2}}{2\mu}
+\frac{\log 2}{\mu \xi} \nonumber\\
&=&\frac{\cos s \sin s}{\mu^2}\left(\cot^2 s \log|\sec s|+\log 2 \right),\nonumber
\end{eqnarray}
where use has been made of the Lemma 2 in Appendix C,
$H \left[\frac{1}{\xi}\right]=-\pi \delta(\cdot)$ in the second line,
$H [\delta ]=   \frac{1}{\pi\xi}$ in the third
and $s=\tan^{-1}\frac{\xi}{\mu}$ in the fourth.
Hence we find
$$K[s\cot s]=\cos s \sin s \left(\cot^2 s \log|\sec s| +\log 2\right),$$
that is,
$$z'(s) \approx s \cot s  +\cot^2 s\log|\sec s| +\log 2.$$
We refrain from comparing the second-order approximation with the solution obtained by
the Newton-Raphson method (Section 3.4) as it is  difficult to fix the prefactor.

\subsection{Determination of the self-similar profile by the Newton-Raphson method}
As it is difficult to solve the determining equation (\ref{deteq_hypo2}) analytically,
we solve it numerically using the Newton-Raphson iteration method.
By integrating (\ref{steadyNL1}) and dropping an integration constant
under the boundary conditions $U(\xi) \to 0$ as $|\xi| \to \infty,$  we find
$$UU_\xi=-\nu'\Lambda U+a\xi U_{\xi},$$
where
$\Lambda U=\frac{1}{\pi}{\rm p.v.}\int_{-\infty}^{\infty}\frac{U(\xi)-U(\eta)}{(\xi-\eta)^2}d\eta.$
\textcolor{black}{See Appendix A.2 for the details of th numerical scheme.}

We take the following parameters 
$\nu=4,a=1,L=20,N=1000, M=1$ for the  numerical computations.
Starting from the linear solution
$U^{(0)}(\xi)=\frac{M}{\pi}\tan^{-1}\frac{\xi}{\mu}$
as an initial guess which corresponds to
$W^{(0)}(\xi)=\frac{M}{\pi} \frac{\mu}{\xi^2+\mu^2},$ 
 it converges very quickly, say, just after 10 iterations.

\begin{figure}
\begin{minipage}{.7\linewidth}\
  \includegraphics[scale=.7,angle=0]{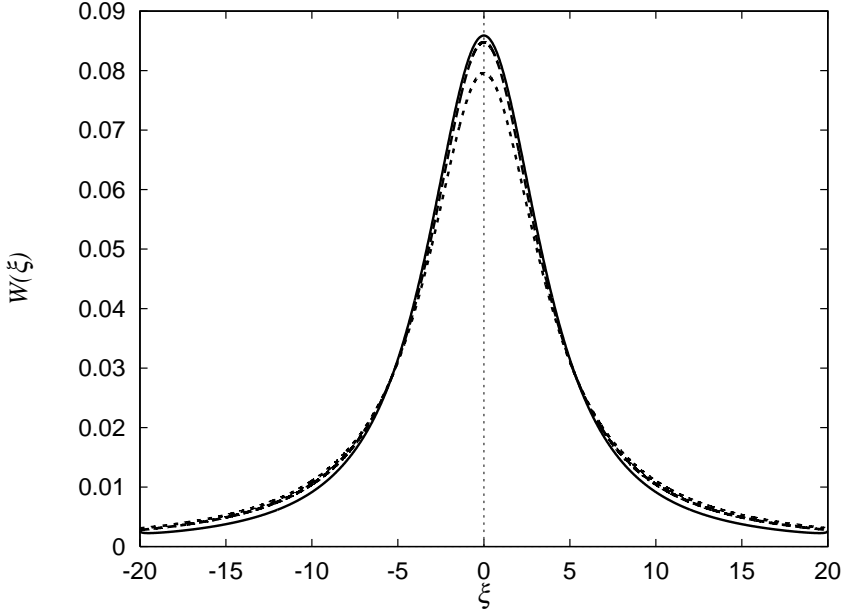}
  \caption{Self-similar profile $W(\xi)$ of the hypoviscous Burgers equation in velocity gradient;
    the exact solution (solid), the first-order approximation (dashed) and the zeroth-order
    approximation (dotted).}
  \label{Fig.2}
\end{minipage}  
\end{figure}   

We regard thus-obtained solution  by the Newton-Raphson iterations   as the exact solution. 
  In Fig.\ref{Fig.2}
  we compare the exact solution (solid)  with the first-order approximation (\ref{W_1st}) to the Poincar{\'e}'s
  variational equation (dashed),
  \textcolor{black}{where the Reynolds number is $R=\frac{M}{\nu}=\frac{1}{4}$.}
  The zeroth-order linear solution, the Cauchy kernel, is also depicted for comparison (dotted).
  Near the origin for small $|\xi|$, the first-order approximation shows an excellent agreement with the exact
  solution, while for larger $|\xi|$, the linear solution shows a better agreement with the exact solution.
  \textcolor{black}{We have confirmed that the first-order approximation improves for smaller $R$, by changing
    $R=1, 1/2,$ and $1/4.$}
  We conclude that overall the  first-order approximation captures the core part of the profile successfully,
  despite the rather crude nature of approximations adopted.
  
\subsection{Lifting the self-similar solution to a more general one}
Assuming that $f(s)$ is obtained, we explain how we can construct more general solutions.

           {\bf Step 1.} Assume that $U(\xi)=Cf(s),\; s(=\widehat{U})\equiv \tan^{-1}\frac{\xi}{\mu}$
           and we write
$$W(\xi)=C f'(s)\frac{\mu}{\xi^2+\mu^2} =  F( \widehat{U}(\xi);C\partial_\xi \widehat{U}(\xi)),$$
           where  $W(\xi)$ is a \textit{near-identity} transformation of $C\partial_\xi \widehat{U}(\xi)$.
Here $F$ satisfies the scaling $F(x;\alpha y)=\alpha F(x;y)$ for $\forall \alpha >0$.           
Reverting to the original variables, a particular self-similar solution is obtained as
$$w(x,t)=\frac{1}{a(t+t_*)}                                                              
F\left(\widehat{U}\left(\frac{x}{a(t+t_*)}\right);C\partial_\xi \widehat{U}\left(\frac{x}{a(t+t_*)}\right)\right),$$
where $at_*=1.$

{\bf Step 2.} Replacing the self-similar Poisson flow with the general   Poisson flow, we find
\begin{eqnarray}
  w(x,t) %&=&\frac{1}{a(t+t_*)} W(\xi,\tau) \nonumber\\
&=&\frac{1}{a(t+t_*)} F\left(\widehat{u}\left(x,t\right);C a(t+t_*)\partial_x \widehat{u}\left(x,t\right)\right) \nonumber\\
  &=&C F\left(\widehat{u}\left(x,t\right);\partial_x \widehat{u}\left(x,t\right)\right)
  = C f'(\widehat{u}\left(x,t\right))\,\partial_x \widehat{u}\left(x,t\right),\nonumber
\end{eqnarray}
where $$\widehat{u}(x,t)=\frac{\nu' t}{\pi}\int_{-\infty}^{\infty} \frac{u_0(y)dy}{(x-y)^2+(\nu' t)^2}$$
denotes the general solution to the linearised equation  by the Poisson kernel.

An analogue of 'Cole-Hopf transformation' for the hypoviscous Burgers equation
$$ w(x,t)= g(\widehat{u}\left(x,t\right))\,\partial_x \widehat{u}\left(x,t\right),$$
where $g(s)\equiv f'(s)$ is determined {\it in principle} by (\ref{deteq_hypo2}).

As it stands, we know the existence of the source-type solution by (\ref{deteq_hypo2}) and
the shape obtained by the Newton-Raphson method (\ref{Fig.2}). We only know the approximate functional
form (\ref{W_1st}), which is too crude for that purpose. Once a precise form of $g(s)$  is determined,
we would be able to integrate the hypoviscous Burgers equation, as the lifting procedure yields
a candidate of the general solutions.

\section{SQG equations}
We discuss the SQG equations in this section. While it is straightforward to obtain the
self-similar solutions under radial-symmetry, it makes sense to write them down explicitly in
various dependent variables.
\subsection{SQG equation with standard dissipativity}
In two dimensions the incompressible velocity field $\bm{u}$ is derived from the stream function $\psi$.
Let $\bm{u}=\nabla^\perp \psi$ and the active scalar $\theta$ is related to it through the constitutive equation $\theta=\Lambda \psi,$
where $\Lambda \equiv(-\triangle)^{1/2}$ denotes the Zygmund operator.
The SQG equation with  standard dissipativity can be written as
\bel{theta.eq}
\frac{\partial \theta}{\partial t}+\nabla \cdot(\theta\bm{u})=\nu \triangle \theta.
\ee
Type 1 critical scale-invariance is achieved with $\psi$. The governing equation for $\psi$ reads
\bel{psi.eq} 
\frac{\partial \psi}{\partial t}-\bm{R}\cdot\left((\nabla^{\perp}\psi)\Lambda \psi \right)=
\nu \triangle \psi,
\ee
which can be derived by applying $\Lambda^{-1}$ to (\ref{theta.eq}).
Here $\bm{R}$ stands for the Riesz transform defined by
$$R_i[f](\bm{x})=-\partial_i \Lambda^{-1} f(\bm{x})\\
=\frac{1}{2\pi}{\rm p.v.}\int_{\mathbb{R}^2}\frac{y_i}{|\bm{y}|}f(\bm{x}-\bm{y})d\bm{y},\;
(i=1,2).$$
Making (\ref{psi.eq})  explicit, we can equivalently write it as
\bel{psi.eq2}
\frac{\partial \psi}{\partial t}=\frac{1}{2\pi}{\rm p.v.}\int_{\mathbb{R}^2}
\frac{(\bm{x}-\bm{y})\times\nabla\psi(\bm{y})\Lambda \psi(\bm{y})}{|\bm{x}-\bm{y}|^3} dy
+\nu \triangle \psi.
\ee
Applying dynamic scaling transformations
$$\bm{\xi}=\frac{\bm{x}}{\lambda(t)},\tau=\frac{1}{2a}\log \frac{t+t_*}{t_*},$$
$$\theta(\bm{x},t)=\frac{1}{\sqrt{2a(t+t_*)}}\Theta(\bm{\xi},\tau),\;\;\lambda(t)=\sqrt{2a(t+t_*)},\;(2at_*=1)$$
to (\ref{theta.eq}), we have
$$\frac{\partial \Theta}{\partial \tau}+\bm{U}\cdot \nabla\Theta
=\nu \triangle \Theta +a\left(\bm{\xi}\cdot \nabla\Theta+\Theta\right).$$
Type 2 critical scale-invariance is achieved with the gradient of $\Theta$,
(i.e. the second derivatives of $\Psi$)
in the sense that its governing equation acquires the Fokker-Planck operator:
$$\frac{\partial \nabla \Theta}{\partial \tau}
  +(\bm{U}\cdot \nabla) \nabla\Theta +(\nabla\bm{U})^T \nabla\Theta
  =\nu \triangle \nabla \Theta+a\nabla\cdot(\bm{\xi}\otimes \nabla\Theta),$$
where $T$ denotes matrix transpose.  
%$$\partial_j(U_j \partial_i \Theta +y \partial_i U_j\, \Theta)
%=\partial_j (\nu \partial_j  \partial_i \Theta +a \xi_j \partial_i \Theta)$$

For radially-symmetric solutions the advection term vanishes identically and 
the governing equation reads
in the original variables
$$\frac{\partial \theta}{\partial t}=\nu \triangle \theta,$$
where
$\triangle \equiv \frac{1}{r}\frac{\partial}{\partial r} \left(r \frac{\partial}{\partial r}\right)$
denotes the Laplacian in spherical polar coordinates. Remember that $\theta$ has the same physical dimension as
velocity.
After dynamic scaling, the governing equation for $\Theta$ becomes
$$\frac{\partial \Theta}{\partial \tau}=\nu \triangle  \Theta
+a\left(\xi \frac{\partial \Theta}{\partial \xi}+\Theta\right).$$
Whilst $\nabla \theta$ satisfies type 2 scale-invariance, it is more convenient to choose the following
form of the derivative.
That is, in view of the relationship $\omega(r)=\frac{1}{r}\frac{\partial}{\partial r}\left(r u(r)\right)$
between the vorticity $\omega(r)$ and the azimuthal velocity $u(r)$ for the 2D Euler equations,
we are led to introduce $D\Theta \equiv\frac{1}{\xi}\frac{\partial}{\partial\xi}(\xi\Theta).$
We then find 
$$\frac{\partial}{\partial \tau} D\Theta
=\nu \triangle  D\Theta+\frac{a}{\xi}
\frac{\partial}{\partial \xi} \left(\xi^2 D\Theta \right)$$
as its governing equation.
Its steady solution is readily given by
$$D\Theta =C \exp\left(-\frac{a \xi^2}{2\nu} \right),$$
where $C$ denotes a constant. This can be regarded as an analogue of the Burgers vortex
for the 2D Navier-Stokes equations.

\subsection{SQG equation with  hypoviscous dissipativity}
\textcolor{black}{This is the case where  dissipation  comes from the physical mechanism of
  Ekman pumping.}
  The governing equation for the \textcolor{black}{active scalar} $\theta$  reads 
\bel{SQG_hypo}
\frac{\partial \theta}{\partial t}+\bm{u}\cdot \nabla\theta=-\nu' \Lambda \theta,
\ee
where it satisfies type 1 critical scale-invariance.
After applying dynamic scaling transformations
$$\bm{\xi}=\frac{\bm{x}}{\lambda(t)},\tau=\frac{1}{a}\log\frac{t+t_*}{t_*},$$
$$\theta(\bm{x},t)=\Theta(\bm{\xi},\tau),\;\;\lambda(t)=a(t+t_*),\;(at_*=1),$$
we find
$$\frac{\partial \Theta}{\partial \tau}+\bm{U}\cdot \nabla\Theta
=-\nu' \Lambda \Theta +a \bm{\xi}\cdot \nabla\Theta.$$
The type 2 critical scale-invariance is achieved with the second derivative of  $\theta$.
For example, $\triangle \theta$ satisfies after dynamic scaling
\bel{LapThetaEq}
\frac{\partial \triangle\Theta}{\partial \tau}+\bm{U}\cdot \nabla \triangle\Theta
+\triangle\bm{U}\cdot\nabla\Theta+2\nabla \bm{U}:(\nabla\otimes\nabla)\Theta+\bm{U}\cdot\nabla\triangle\Theta\\
=-\nu' \Lambda\triangle  \Theta+a\nabla\cdot(\bm{\xi}\otimes \triangle\Theta).
\ee
If the solution is radial, the left-hand side of (\ref{LapThetaEq}) simplifies due to vanishing
of the nonlinear terms. 
Consider a radial-symmetric solution in spherical polar coordinates, which obey
$$\frac{\partial \theta}{\partial t}=-\nu' \Lambda \theta$$
and after dynamic-scaling it obeys
$$\frac{\partial \Theta}{\partial \tau}=-\nu' \Lambda  \Theta
+a\xi \frac{\partial \Theta}{\partial \xi},$$
where $\xi=|\bm{\xi}|.$
Noting
$$\triangle\left( \xi \frac{\partial \Theta}{\partial \xi}\right)=
\frac{1}{\xi}\frac{\partial}{\partial\xi} \xi\frac{\partial}{\partial\xi}
\left( \xi \frac{\partial \Theta}{\partial \xi}\right)=
\frac{1}{\xi}\frac{\partial}{\partial\xi} \left(\xi^2 \triangle \Theta \right),$$
the governing equation for $\triangle\Theta$ reads
$$\frac{\partial}{\partial \tau}\triangle\Theta=-\nu' \Lambda \triangle  \Theta+\frac{a}{\xi}
\frac{\partial}{\partial \xi} \left(\xi^2 \triangle\Theta \right),$$
and its steady equation becomes
$$\mu \Lambda \triangle \Theta =\frac{1}{\xi}\frac{\partial}{\partial \xi} \left(\xi^2 \triangle \Theta \right),
\;\;(\mu=\nu'/a).$$
Its  solution is given by
$$\triangle\Theta =C\frac{\mu}{(\xi^2+\mu^2)^{3/2}},$$
where $C$ is a constant that depends on $M=\int \triangle\Theta d\bm{\xi}.$
This can be verified by $\Lambda=\nabla\cdot\bm{R}$ and a property of the Riesz transform
\bel{formula}
R_j\left[\frac{\mu}{(\xi^2+\mu^2)^{3/2}} \right]=\frac{\xi_j}{(\xi^2+\mu^2)^{3/2}}\;\;(\mu >0).
\ee
Applying $\partial_j$
$$
\mu\Lambda\left[\frac{1}{(\xi^2+\mu^2)^{3/2}} \right]=\partial_j\frac{\xi_j}{(\xi^2+\mu^2)^{3/2}}
=\frac{2\mu^2-\xi^2}{(\xi^2+\mu^2)^{5/2}}
$$
$$=\frac{1}{\xi}\frac{d}{d\xi} \frac{\xi^2}{(\xi^2+\mu^2)^{3/2}}.$$
%However, because
%c$$\frac{1}{\xi}\frac{d}{d\xi} \frac{\xi^2}{(\xi^2+\mu^2)^{3/2}}
%=\frac{2\mu^2-\xi^2}{(\xi^2+\mu^2)^{5/2}},$$ 
%we find
%$$\mu\Lambda\left[\frac{1}{(\xi^2+\mu^2)^{3/2}} \right]=\frac{1}{\xi}\frac{d}{d\xi} \frac{\xi^2}{(\xi^2+\mu^2)^{3/2}},$$
Hence $\triangle \Theta \propto \frac{1}{(\xi^2+\mu^2)^{3/2}}$.

Alternatively, to confirm the same result we may  start from the explicit form of the linear solution
$$\Theta(\bm{\xi},\tau)=\frac{\nu'}{2\pi}\frac{1-e^{-a\tau}}{a}
\int_{\mathbb{R}^2}
\frac{\theta_0(e^{a\tau}\bm{\eta}) d\bm{\eta}}
{\left\{ |\bm{\xi}-\bm{\eta}|^2  +\nu'^2\left(\frac{1-e^{-a\tau}}{a} \right)^2 \right\}^{3/2}}$$
to deduce
$$\triangle \Theta(\bm{\xi},\tau)=\frac{\nu'}{2\pi}\frac{1-e^{-a\tau}}{a}
\int_{\mathbb{R}^2}
\frac{e^{2a\tau} \triangle_{\eta}\theta_0(e^{a\tau}\bm{\eta}) d\bm{\eta}}
{\left\{ |\bm{\xi}-\bm{\eta}|^2  +\nu'^2\left(\frac{1-e^{-a\tau}}{a} \right)^2 \right\}^{3/2}}$$
$$\to \frac{M}{2\pi}\frac{\mu}{(\xi^2+\mu^2)^{3/2}}\;\mbox{as}\;\tau \to \infty.$$

We wrap up this section by determining functional forms of other variables of physical interest.
Hereafter we take $\mu=1,\;M=1$ for simplicity.
By $\triangle \equiv \frac{1}{\xi}\frac{\partial}{\partial \xi} \left(\xi \frac{\partial}{\partial \xi}\right),$
$\Theta$ satisfies
$$\frac{1}{\xi}\frac{d}{d\xi} \left(\xi \frac{d \Theta(\xi)}{d\xi}\right)
=\frac{1}{2\pi} \frac{1}{(\xi^2+1)^{3/2}},$$
which is integrated to give
$$\Theta=\frac{1}{2\pi}\log\left(1+\sqrt{\xi^2+1}\right).$$
We note that the constitutive equation $\Omega=\Lambda \Theta,$ or equivalently,
$$\Lambda \Omega= -\triangle \Theta$$
holds for both SQG equations with standard and hypo dissipativity
\textcolor{black}{as it is a kinematic constraint.}
For the latter, the Zygmund operator $\Lambda$ nicely matches self-similar solution in $\triangle \Theta$
of the form of the Poisson kernel.  As a result we can explicitly write the self-similar solution
in vorticity. Actually, applying $R_j=-\Lambda^{-1}\partial_j$ to the above equation, we find
$$R_j[\triangle \Theta]=\partial_j \Omega.$$
Combining this with (\ref{formula}),
we obtain
$$\Omega(\xi)=-\frac{1}{2\pi }\frac{1}{(\xi^2+1)^{1/2}}.$$
As a by-product, the above results yield formulas for some definite integrals of interest,
see Appendix D.

We summarise self-similar profiles of the 2D fluid equations and the SQG equation 
in Table 2, including the expressions for velocity.
\textcolor{black}{We note that for the  2D fluid equations, with $\triangle$ the dynamically scaled vorticity $\Omega$ coincides with the Burgers vortex,
whereas  with  $\Lambda$ it is $\triangle U$ that coincides with the hypoviscous Burgers vortex $\omega \sim\frac{1}{(r^2+1)^{3/2}}$.}
\begin{table}[h]
\begin{center}     
  \caption{Suitably normalised self-similar solutions of 2D incompressible and SQG equations.}
\begin{tabular}{ccc}\hline                                                      
dissipativity  & 2D incompressible & SQG \\ \hline
Laplacian operator $\triangle$ \\(with $\nu/a=2$) & $\begin{array}{c} \Omega\equiv-\triangle \Psi\\
  =\frac{1}{\pi} e^{-\xi^2}\end{array}$ &
$\begin{array}{c} D \Theta\equiv\frac{1}{\xi}\partial_{\xi}(\xi\Theta)\\
  =\frac{1}{\pi}  e^{-\xi^2}\end{array}$ \\ \noalign{\vskip 0.2cm}\hline \noalign{\vskip 0.3cm}
Zygmund operator $\Lambda$\\ (with $\nu'/a=1$)& $\begin{array}{c} \triangle U =\frac{1}{2\pi} \frac{1}{(\xi^2+1)^{3/2}} \\  \noalign{\vskip 0.3cm}
   U =\frac{1}{2\pi} {\small \log(1+\sqrt{\xi^2+1})} \\ \noalign{\vskip 0.3cm}
\Omega = \frac{\{\xi \log(1+\sqrt{\xi^2+1}) \}'}{2\pi \xi}\end{array}$ &
$\begin{array}{c} \triangle \Theta =\frac{1}{2\pi} \frac{1}{(\xi^2+1)^{3/2}} \\  \noalign{\vskip 0.3cm}
  \Theta =\frac{1}{2\pi} {\small \log(1+\sqrt{\xi^2+1})} \\  \noalign{\vskip 0.3cm}
  U =-\frac{1}{2\pi} \frac{\xi}{1+\sqrt{\xi^2+1}} \\  \noalign{\vskip 0.3cm}
  \Omega =-\frac{1}{2\pi} \frac{1}{(\xi^2+1)^{1/2}}\end{array}$ \\ \noalign{\vskip 0.3cm} \hline
% \noalign{\vskip 0.3cm}
\end{tabular}
\end{center}
\label{2D-SQG}
\end{table}      

\section{Summary and outlook}
\color{black}
A method of studying nonlinear PDEs on the basis of the 'heat' solutions is identified
and tested on the hypoviscosity versions of the SQG equation and, as a simpler 1D model,
the Burgers equation. The former represents a geophysical system of strongly stratified fluids.

After reviewing  self-similar solutions of the standard Burgers equation,
we  studied those of the hypoviscous Burgers equation.
For the standard Burgers equation the crucial equation that determines
the near-identity transformation underlying the Cole-Hopf linearisation is
as simple as (\ref{deteq_standard}).
For the hypoviscous Burgers equation we have successfully derived the determining equation
(\ref{deteq_hypo2}) for the near-identity transformation, the very existence of which indicates
integrability of the hypoviscous Burgers equation by the Poisson  kernel.
Its systematic approximation is formalised and worked out to the first-order.
We have shown numerically that it well captures the nonlinear self-similar profile that is
obtained by the Newton-Raphson scheme.
Extension to more general dissipativity $\Lambda^{\alpha}\;\;(0 < \alpha < 2)$ seems challenging,
because useful formulas for handling this case are unavailable and
an explicit expression for the Green's function is lacking. 

We also studied self-similar solution to the 2D incompressible fluid equations and the SQG equations,
with both standard and hypoviscosity. We sorted out and tabulated the self-similar profiles in various
dependent variables.

It is of interest to extend the current method and apply it to more general non-integrable equations,
such as the 3D Navier-Stokes equations. In that case we know the linear solutions explicitly \cite{OV2022}
and that a trace of the nonlinear terms {\it  do} remain in the self-similar profiles.
By studying the trace (at least) numerically,
it is worthwhile for us to see what we can do about it and see if there is anything we can learn from it.
This will be left for future research.
\color{black}

%subject

%scope

%purpose

%organization

\appendix
\color{black}
\section{Newton-Raphson method}
\subsection{Burgers equation}
Introduce  a discretisation
$$U_i=U(\xi_i),\; \xi_i=h\,i,\;\; (i=-N,\ldots,N),$$
where $h=L/N$ denotes the mesh size, $L$ the size of the interval and $N$ the number of grid points.
Defining
$$J_i(U) \equiv \nu \frac{U_{i+1}-U_{i-1}}{2h}+a \xi_i U_i-\frac{U_i^2}{2},\;\;
(i=-N,\ldots,N),$$
the equation  (\ref{steady_sc_Burg}) becomes $J_i(U)=0, \;\;(i=-N,\ldots,N).$
We have
$$\frac{\partial J_i}{\partial U_j}=A\delta_{i+1,j} - A\delta_{i-1,j}+B_i \delta_{i,j},
(i,j=-N,\ldots,N),$$
where $A\equiv\frac{\nu}{2h}, B_i\equiv a\xi_i -U_i$
and $\delta_{i,j}$ denotes Kroneckers's delta.
By Taylor-expanding the function $J$, we find
$$J_i(U+\delta U)=J_i(U)+\frac{\partial J_i}{\partial U_j}\delta U_j.$$
Setting $J_i(U+\delta U)=0,$  we have
$$\frac{\partial J_i}{\partial U_j}\delta U_j=-J_i(U),$$
and the iteration scheme comes out as
%$$\widetilde{U}_{i}=U_{i}-\left(\frac{\partial J_i}{\partial U_j}\right)^{-1}J_j(U)$$
$$U^{(n+1)}_i=U^{(n)}_{i}-\left(\frac{\partial J_i(U^{(n)})}{\partial U^{(n)}_j}\right)^{-1}J_j(U^{(n)})$$
for $i=-N,\ldots,N,$  where $U^{(n)}_{i}$ denotes the $n$-th iterate ($n=0,1,\ldots$).

\subsection{SQG equation}
We define
$$J(U)\equiv -\nu' \frac{1}{\pi}{\rm p.v.}\int_{-\infty}^{\infty}\frac{U(\xi)-U(\eta)}{(\xi-\eta)^2}d\eta
+a\xi U_{\xi}-UU_{\xi}$$
and introduce the corresponding discretisation
$$J_i(U)\equiv - \frac{\nu' h}{\pi} \sum_{k=-N}^{N}\frac{U_i-U_k}{(\xi_i-\xi_k)^2}
+a\xi_i\frac{U_{i+1}-U_{i-1}}{2h}-U_i\frac{U_{i+1}-U_{i-1}}{2h},$$
where $\xi_i=h\,i\; h=L/N,\; i=-N,\ldots,N,$
with suitable handling at end points,
e.g. $\frac{U_{N+1}-U_{N-1}}{2h} \to \frac{U_{N}-U_{N-1}}{h}$.
The Jacobian is given by
$$\frac{\partial J_i}{\partial U_j}
= - \frac{\nu' h}{\pi}\left\{\delta_{i,j} \sum_{k=-N}^{N}\frac{1}{(\xi_i-\xi_k)^2}
-\frac{1}{(\xi_i-\xi_j)^2}\right\}
-\delta_{i,j}\frac{U_{i+1}-U_{i-1}}{2h}
  +\delta_{i+1,j}\frac{a \xi_i-U_i}{2h}  -\delta_{i-1,j}\frac{a \xi_i-U_i}{2h}.$$
  As before, setting
  $$0=J_i(U+\delta U)=J_i(U)+\frac{\partial J_i}{\partial U_j}\delta U_j,\;\; i=-N,\ldots,N,$$
  we obtain the iteration scheme as follows:
  $$U^{(n+1)}_{i}=U^{(n)}_{i}-\left(\frac{\partial J_i(U^{(n)})}{\partial U^{(n)}_j}\right)^{-1}J_j(U^{(n)})$$
for $i=-N,\ldots,N,$  where $U^{(n)}_{i}$ denotes the $n$-th iterate, ($n=0,1,\ldots$).
We carried out iterations numerically using the $L$-$U$ decomposition.
\color{black}
\section{Lemma 1}
We verify
  $$\int_0^s\log (\sin r) dr=-s \log 2- \frac{1}{2}\sum_{n=1}^{\infty}\frac{\sin(2ns)}{n^2}.$$
{\bf Derivation}\\
Writing $\int_0^s\log(\sin r) dr=s\log (\sin s)-\int_0^s r\frac{\cos r}{\sin r}dr,$
  we have
  $$I(s)=s\log (\sin s)-\int_0^s\log (\sin r) dr.$$
  Recalling  the following identity
%  $$\sum_{n=1}^{\infty}\frac{\sin(n\theta)}{n}=\frac{\pi-\theta}{2},$$
  $$\sum_{n=1}^{\infty}\frac{\cos(n\theta)}{n}=-\log\left(2\sin\frac{\theta}{2}\right)\;\;
\mbox{for}\;\; 0 < \theta <2 \pi$$
and integrating from  0 to $s$ with respect to $\theta$, we have
  $$-\sum_{n=1}^{\infty}\frac{\sin(n s)}{n^2}=\int_0^s \log\left(2\sin\frac{\theta}{2}\right)d\theta$$
  $$=s \log 2+2\int_0^{s/2} \log (\sin r) dr.$$
  Letting $s \to 2s,$ we find
  %  $$-\sum_{n=1}^{\infty}\frac{\sin(2ns)}{n^2}=2s \log 2+2\int_0^{s} \log (\sin r) dr.$$
  $$I(s)=s\log(2\sin s)+2\sum_{n=1}^{\infty}\frac{\sin(2ns)}{(2n)^2},$$
  hence the desired result follows. $\square$

  It should be noted that the integral $I(s)$ cannot be represented in terms of
  elementary functions.   It can be expressed, however, by the dilogarithm
  function defined by
  ${\rm Li}_2(z)=\sum_{n=1}^{\infty} \frac{z^n}{n^2}\;\mbox{for}\;|z| \le 1.$
In fact, it can be readily verified that
  $$I(s)=s\log(2\sin s)+\frac{1}{2} \Im\left({\rm Li}_2(e^{2 i s}) \right),$$
  where $\Im$ denotes the imaginary part.

\section{Lemma 2}
We verify
  $$H\left[ \frac{\tan^{-1}\frac{\xi}{\mu}}{\xi^2+\mu^2}\right]
=\frac{1}{\xi^2+\mu^2}\log\frac{\sqrt{\xi^2+\mu^2}}{2\mu},\;(\mu >0).$$

{\bf Derivation}\\
Clearly it is sufficient to show that
$$H\left[ \frac{\tan^{-1} x}{x^2+1}\right]=\frac{1}{x^2+1}\log\frac{\sqrt{x^2+1}}{2}.$$
Put
$$f(\lambda)= H\left[ \frac{\tan^{-1}\lambda x}{x^2+1}\right],$$
and we have
$$f'(\lambda)= H\left[ \frac{1}{x^2+1}\frac{x}{1+(\lambda x)^2} \right]
=\frac{1}{\lambda^2} H\left[ \frac{1}{x^2+1}\frac{x}{x^2+ 1/\lambda^2} \right].$$
By
$$H\left[ \frac{x}{(x^2+a^2)(x^2+b^2)} \right]=\frac{x^2-ab}{(a+b)(x^2+a^2)(x^2+b^2)},\;(a,b >0),$$
cf. \cite{King2009}, we have
$$f'(\lambda)=\frac{\lambda x^2-1}{(\lambda+1)(x^2+1)(\lambda^2 x^2+1)}
=\frac{1}{x^2+1} \left(\frac{x^2 \lambda }{x^2 \lambda^2 +1}-\frac{1}{\lambda+1} \right).$$
Integrating with respect to $\lambda$ we get 
$$f(\lambda)=\frac{1}{x^2+1}\log \frac{\sqrt{x^2 \lambda^2 +1}}{\lambda+1}.$$
Setting $\lambda=1$ we obtain the desired result. $\square$

\section{Some definite integrals}
Carrying out angular integration we can confirm, for radially symmetric functions $f(\bm{x})=f(r),\;r\equiv|\bm{x}|$,
  \begin{eqnarray}
    \Lambda^{-1}f(\bm{x})&=&\frac{1}{2\pi}\int_{\mathbb{R}^2} \frac{f(\bm{y})}{|\bm{x}-\bm{y}|}d\bm{y} \nonumber\\
    &=& \frac{1}{2\pi} \int_0^\infty \int_0^{2\pi}\frac{f'(r') r'dr' d\theta}{(r^2+r'^2-2rr'\cos\theta)^{1/2}},  \nonumber\\
    &=& \frac{r}{2\pi}\int_0^\infty ds s f(rs) \int_0^{2\pi}\frac{d\theta}{(1+s^2-2s\cos \theta)^{1/2}},
    \;\mbox{by}\;s=r'/r\nonumber\\
    &=& \frac{2}{\pi}\int_0^\infty \frac{sf(s)}{r+s}K\left(\frac{2\sqrt{rs}}{r+s} \right)ds, \;\mbox{after renaming}\;r'\to s. \nonumber    
 \end{eqnarray}
Likewise we have
  \begin{eqnarray}
  \Lambda f(\bm{x})&=&\frac{1}{2\pi}
  {\rm p.v.}\int_{\mathbb{R}^2} \frac{f(\bm{x})-f(\bm{y})}{|\bm{x}-\bm{y}|^3}{\rm d}\bm{y}\nonumber\\
  &=&\frac{2}{\pi} {\rm p.v.}\int_{0}^{\infty} \frac{f(r)-f(s)}{(r+s)(r-s)^2}E\left( \frac{2\sqrt{rs}}{r+s}\right)s ds.\nonumber
  \end{eqnarray}
  Here use has been made of the following formulas, e.g. \cite{GR2007}, p.443
  $$\int_0^\pi \frac{d\theta}{(1-2a\cos\theta+a^2)^{1/2}}
    =\frac{2}{a+1}K\left( \frac{2\sqrt{a}}{a+1}\right),\;(a >0),$$          
  $$\int_0^\pi \frac{d\theta}{(1-2a\cos\theta+a^2)^{3/2}}
  =\frac{2}{(a+1)(a-1)^2}E\left( \frac{2\sqrt{a}}{a+1}\right),\;(a > 0),$$
  where $E(k)=\int_0^{\pi/2}\sqrt{1-k^2 \sin^2 u} du$ denotes the complete elliptic integral of the first kind
  and $K(k)=\int_0^{\pi/2}\frac{du}{\sqrt{1-k^2 \sin^2 u}}$ that of the second kind. 

It is of interest to observe that the following formulas hold for the two definite integrals.
It follows from $\Lambda^{-1}\triangle\Theta=-\Omega$ that
$$\frac{2}{\pi}\int_0^\infty \frac{s}{r+s}\,\frac{1}{(s^2+1)^{3/2}}K\left(\frac{2\sqrt{rs}}{r+s} \right)ds
=\frac{1}{(r^2+1)^{1/2}}.$$
It also follows from $\Lambda\Theta=\Omega$ that
$$\frac{2}{\pi}{\rm p.v.}\int_{0}^{\infty} \frac{1}{(r+s)(r-s)^2}\,
{\small\log\left(\frac{1+\sqrt{r^2+1}}{1+\sqrt{s^2+1}}\right)}
E\left(\frac{2\sqrt{rs}}{r+s} \right)ds =\frac{-1}{(r^2+1)^{1/2}}.$$
%$$\Omega(\xi)=\frac{a}{2\pi\nu} e^{-\frac{a\xi^2}{2\nu}},\;U(\xi)=\frac{1}{2\pi \xi}(1- e^{-\frac{a\xi^2}{2\nu}}),\;
%\Psi(\xi)=\frac{1}{4\pi}\left({\rm Ei}\left(-\frac{a\xi^2}{2\nu}\right)-\log \xi^2\right)$$

\ack
This work was supported by the Research Institute for Mathematical
Sciences, an International Joint Usage/Research Center located in Kyoto
University. This work was also supported by JSPS KAKENHI Grant Number JP21K20322.

\end{document}